\documentclass[acmsmall]{acmart}

\AtBeginDocument{%
  \providecommand\BibTeX{{%
    \normalfont B\kern-0.5em{\scshape i\kern-0.25em b}\kern-0.8em\TeX}}}

\setcopyright{rightsretained}
\acmPrice{}
\acmDOI{10.1145/3622858}
\acmYear{2023}
\copyrightyear{2023}
\acmSubmissionID{oopslab23main-p471-p}
\acmJournal{PACMPL}
\acmVolume{7}
\acmNumber{OOPSLA2}
\acmArticle{282}
\acmMonth{10}
\received{2023-04-14}
\received[accepted]{2023-08-27}

\usepackage{url}
\usepackage[english]{babel}
\usepackage[utf8]{inputenc}
\usepackage{algorithm}
\usepackage[noend]{algpseudocode}
\usepackage{fancyvrb}
\usepackage{xspace}
\usepackage{mathpartir}
\usepackage{hyperref}
\usepackage{paralist}
\usepackage{adjustbox}
\usepackage{float}
\usepackage{circledsteps}

\usepackage{graphicx}
\usepackage{caption}
\usepackage{subcaption}
\usepackage{setspace}

\usepackage{tabularx}

\newcommand{\zph}{\phantom{0}}
\newcommand{\zzph}{\phantom{0}}

\newif\ifanonymous
\anonymoustrue

\usepackage{fancyvrb}
\RecustomVerbatimEnvironment{Verbatim}{Verbatim}{fontsize=\smaller}

\usepackage{listings}
\newcommand*\LSTfont{\fontencoding{T1}\ttfamily\SetTracking{encoding=*}{-60}\lsstyle}
\usepackage{cleveref}
\usepackage{tikz}

\newcommand{\todo}[1]{}

\def\|#1|{\mathid{#1}}
\newcommand{\mathid}[1]{\ensuremath{\mathit{#1}}}
\def\<#1>{\codeid{#1}}
\protected\def\codeid#1{\ifmmode{\mbox{\smaller\ttfamily{#1}}}\else{\smaller\ttfamily #1}\fi}

\newcommand{\CreatesMustCallFor}{@Creates\-Must\-Call\-For}
\newcommand{\EnsuresCalledMethods}{@Calls}
\newcommand{\MustCallAlias}{@Re\-source\-Alias}
\newcommand{\MustCall}{@Must\-Call}
\newcommand{\NotOwning}{@Not\-Own\-ing}

\newcommand{\InstanceField}{\textsc{Field}}
\newcommand{\InstanceMethod}{\textsc{Method}}
\newcommand{\FieldMustCall}{\textsc{Field\-Disposal}}

\begin{document}
\hypersetup{pageanchor=false}

\title{Inference of Resource Management Specifications}

\author{Narges Shadab}
\orcid{0009-0008-4417-9416}
\email{nshad001@ucr.edu}
\affiliation{
  \institution{University of California, Riverside}
  \city{Riverside, CA}
  \country{USA}
}

\author{Pritam Gharat}
\orcid{0000-0002-5961-8142}
\email{t-prgharat@microsoft.com}
\affiliation{
  \institution{Microsoft Research}
  \city{Bengaluru}
  \country{India}
}

\author{Shrey Tiwari}
\orcid{0000-0001-6697-0219}
\email{shreymt@gmail.com}
\affiliation{
  \institution{Microsoft Research}
  \city{Bengaluru}
  \country{India}
}

\author{Michael D. Ernst}
\orcid{0000-0001-9379-277X}
\email{mernst@cs.washington.edu}
\affiliation{
  \institution{University of Washington}
  \city{Seattle, WA}
  \country{USA}
}

\author{Martin Kellogg}
\orcid{0000-0002-3185-2340}
\email{martin.kellogg@njit.edu}
\affiliation{
  \institution{New Jersey Institute of Technology}
  \city{Newark, NJ}
  \country{USA}
}

\author{Shuvendu K. Lahiri}
\orcid{0000-0002-4446-4777}
\email{Shuvendu.Lahiri@microsoft.com}
\affiliation{
  \institution{Microsoft Research}
  \city{Redmond, WA}
  \country{USA}
}

\author{Akash Lal}
\orcid{0009-0002-4359-9378}
\email{akashl@microsoft.com}
\affiliation{
  \institution{Microsoft Research}
  \city{Bengaluru}
  \country{India}
}

\author{Manu Sridharan}
\orcid{0000-0001-7993-302X}
\email{manu@cs.ucr.edu}
\affiliation{
  \institution{University of California, Riverside}
  \city{Riverside, CA}
  \country{USA}
}

\renewcommand{\shortauthors}{Shadab, Gharat, Tiwari, Ernst, Kellogg, Lahiri, Lal, and Sridharan}

\begin{abstract}
A resource leak occurs when a program fails to free some finite resource
after it is no longer needed. Such leaks are a significant cause of
real-world crashes and performance problems.
\ifanonymous Recent work \else We recently \fi
proposed an approach to prevent resource leaks based on checking
\textit{resource management specifications}.
A resource management specification expresses how the program allocates
resources, passes them around, and releases them; it also tracks
the ownership relationship between objects and resources, and
aliasing relationships between objects.
While this specify-and-verify approach has several advantages compared to
prior techniques, the need to manually write annotations presents a
significant barrier to its practical adoption.

This paper presents a novel technique to automatically infer a resource
management specification for a program, broadening the applicability of
specify-and-check verification for resource leaks.  Inference in this domain is
challenging because resource management specifications differ significantly in
nature from the types that most inference techniques target.  Further, for
practical effectiveness, we desire a technique that can infer the resource
management specification intended by the developer, even in cases when the code
does not fully adhere to that specification.  We address these challenges
through a set of inference rules carefully designed to capture real-world coding
patterns, yielding an effective fixed-point-based inference algorithm.

We have implemented our inference algorithm in two different systems,
targeting programs written in Java and C\#.
In an experimental evaluation,
our technique inferred 85.5\%\todo{Let's round this to the nearest percent.}
of the annotations that programmers had written
manually for the benchmarks.
Further, the verifier issued nearly the same rate of false alarms with the
manually-written and automatically-inferred annotations.
\end{abstract}


\maketitle

\section{Introduction}\label{sec:intro}

A resource leak occurs when a finite resource managed by the programmer is
not released after it is no longer needed, causing the resource
to be held indefinitely by the program.
For example, resources such as file descriptors, sockets, and database
connections must be explicitly released by the programmer.
Resource leaks can lead to the depletion of system resources and poor performance, eventually causing the program or the whole system to crash.

\ifanonymous Recent work \else We recently \fi
proposed a specify-and-verify approach to find and prevent resource
leaks~\cite{KelloggSSE2021}.  This work first requires the programmer to
write a \textit{resource management specification} of how the program
intends to manage its resources.
The specification language includes annotations that indicate which objects
control a resource, and lightweight ownership and
aliasing properties that track the flow of resources through the program. 
Given an annotated program,
a verification tool then verifies the correctness of the annotations and
concludes the absence of leaks given the annotations.
The annotations further allow the checker to be fast through modularity and
incrementality: it only needs to analyze one method at a time, and after a
code change, it only needs to analyze modified methods.

However, there is a substantial barrier to practical
adoption of this approach:  the programmer must write the resource management specification.
Understanding and specifying resource management protocols can be very
challenging, especially for large legacy systems.

This paper presents a novel inference technique to automatically discover a
resource management specification that can then be used for
verification.  Automated inference significantly reduces user effort and
broadens the applicability of this style of verification.

Inference of resource management specifications poses multiple challenges.
These specifications differ from the types or
type qualifier properties~\cite{FosterFFA99} that most existing inference
techniques have focused on in the past.  Resource management specifications
must capture multiple inter-related properties, including resource
ownership, which methods release resources, and aliasing relationships.  Effective inference of these properties 
requires a customized algorithm that infers these facts simultaneously.
Furthermore, it is desirable to infer the \emph{intended} specification
because a program may be buggy, such as releasing a resource along some
control-flow paths but not all.  The need for \emph{optimistic inference}
(inferring specifications that cannot be verified)
is atypical for inference
techniques. 

Our algorithm performs inference based on how the program creates, passes
around, and releases resources.
The
inference is bootstrapped from tool-provided annotations for resource types in the standard libraries (the JDK and the .NET
framework) and optional programmer-written annotations.
Our algorithm has two phases.

The first phase determines the owned resources for each class
by identifying fields whose declared type might need to be released. It
also identifies a
``disposal
method''\footnote{This disposal method is often named \<close> (in Java)
or \<Dispose> (in C\#).}
of a class that releases the owned resources.  A method is optimistically inferred to
be a disposal method if it releases a resource on some (but not necessarily all)
paths, thereby capturing what is typically the intended protocol.  With this optimistic technique, the checker will still report an error after inference is completed, but the error is localized to where the likely-intended specification is violated, and hence where the actual code fix is likely to be needed.

The second phase infers all other parts of a resource management specification.
This includes optimistic inference of method signatures (ownership of
method parameters and returns) and inference of resource
alias relationships.
When closing any of multiple objects in the program is
adequate to release an underlying resource, we refer to those objects
as \emph{resource aliases}.

This paper formalizes the inference algorithm as a set of inference rules,
such that annotations are inferred by applying the rules to a
fixed point.  To demonstrate the generality and effectiveness of our
approach, we implemented two different instantiations of the algorithm.
The first implementation is for Java programs, built on the Checker
Framework~\cite{PapiACPE2008} that supports type checking, local type inference,
and abstract interpretation.
The second implementation is for C\# programs, built on
the CodeQL framework~\cite{codeql} that provides a Datalog-like declarative code query 
language for building custom program analyses.

We performed an experimental evaluation on programs that had been manually
annotated and verified (in some cases the verification required suppressing
false positive warnings).  We ran our technique on un-annotated versions of the
programs, and it inferred
most of the hand-written annotations (85.5\%).  Further,
most remaining
verifier warnings are due to true positive issues in the code or verifier
imprecision, not missing or incorrectly-inferred annotations.

\paragraph*{Contributions} This paper makes the following contributions:
\begin{enumerate}
    \item a novel optimistic inference algorithm for resource management
      specifications, designed to infer specifications likely intended by the developer;
    \item a formalization of the algorithm that is generic across programming languages;
    \item implementations for Java and for C\#, demonstrating the generality of our approach;
    \item experiments that show that the approach is effective in practice.
\end{enumerate}

\section{Background: Resource management specifications}

Every program must follow the contract that after a resource is acquired,
the resource must be released,
permitting the runtime or operating system to reuse the resource.
We call the method releasing a resource the ``disposal method.''
In Java, the disposal method is often named \<close>, while in
C\# it is often named \<Dispose>.  It is an obligation on the programmer to
call this method on each object managing a resource after the program is done using it.

\Cref{sec:annotation-syntax} reviews the syntax and semantics for the specifications used by the Resource Leak Checker~\cite{KelloggSSE2021}, via which a programmer
communicates
how the program allocates resources,
passes them around, and releases them.
\Cref{sec:class-specification} gives an example resource management specification using this specification language, and finally \Cref{sec:verification-algorithm} briefly describes modular verification of these specifications.

\subsection{Annotation syntax and semantics} 
\label{sec:annotation-syntax}

A resource management specification is expressed in Java as annotations, which start with
an at-sign (\<@>).  C\# uses attributes instead, which are enclosed within
square brackets (\<[>\ldots\<]>).  This section uses the Java syntax~\cite{KelloggSSE2021}.
There are annotations for expressing required and guaranteed calls, lightweight ownership hints, and resource aliasing.
For additional details, see the Resource Leak Checker manual~\cite{rlc-manual}.

\smallskip\noindent
{\bf \<\MustCall(m)>}
is a \emph{type qualifier} that modifies a type.  The argument \<m> is a
method that must be called on any value of that type.  We call \<m> the
type's \emph{must-call obligation}.  \<\MustCall()> (without an annotation
argument) means that no method is required to be called.
A type qualifier can be written on a type use or a type declaration.

Here is an example of a \<\MustCall> annotation on a type
\emph{use}:\footnote{Our code examples use {\bfseries boldface} for declared
identifiers and {\color{blue}blue} for resource leak specifications.}

\begin{Verbatim}[commandchars=\|\[\]]
[|color[blue]@MustCall("print")] Diagnostic[|bfseries explainToUser]() { ... }
\end{Verbatim}

\noindent
The return type of \<explainToUser> is \<@MustCall("print") Diagnostic>,
where \<@MustCall("print")> is a type qualifier and \<Diagnostic> is the Java basetype.
The program must call \<print> on the returned diagnostic.  This annotation
ensures that a client does not create a diagnostic but forget to print it.

When written on a type \emph{definition}, a \<\MustCall> annotation
indicates a method that must be
called on every object of that type (including subtypes).
For example, the JDK's
\<java.io.Closeable> interface is specified as:

\begin{Verbatim}[commandchars=\|\[\]]
 [|color[blue]@MustCall("close")] // close() must be invoked on every Closeable object
 public interface[|bfseries Closeable] { ... }
\end{Verbatim}

\noindent
This annotation makes writing \<Closeable> equivalent to writing
\<@MustCall("close") Closeable>.  That behavior may be overridden:
here is an annotation on a type \emph{use}
in \<java.io.OutputStream>:

\begin{Verbatim}[commandchars=\|\[\]]
[|color[blue]@MustCall()] OutputStream[|bfseries nullOutputStream]() { ... }
\end{Verbatim}

\noindent
This method returns an \<OutputStream> with no must-call obligation, because
a null output stream need not be closed.

\smallskip\noindent
{\bf \<\EnsuresCalledMethods(exprs, methods)>}
  is a method annotation.  When written on
    a method $m$, it means that $m$ guarantees that all methods in \<methods> are called on
    all expressions referenced in \<exprs>.\footnote{The guarantee only holds when $m$ terminates normally, without throwing an exception.} Here are two examples using \<\EnsuresCalledMethods>:

\begin{Verbatim}[commandchars=\|\[\]]
 // This method reads the first ("#1") formal parameter (x), then closes it.
 [|color[blue]@Calls(exprs = "#1", methods = "close")]
 public int[|bfseries readAndCloseStream](IntStream x) { ... }

 // This method guarantees close() is called on two fields
 [|color[blue]@Calls(exprs={"this.ownedField1", "this.ownedField2"}, methods="close")]
 public void[|bfseries closeFields]() { ... }
\end{Verbatim}

\noindent
\<\EnsuresCalledMethods> enables modular verification in cases where a resource is closed in a callee, e.g.:
\begin{Verbatim}[commandchars=\|\[\]]
  IntStream [|bfseries y] = new IntStream();
  // guaranteed to close y, based on the @Calls annotation
  readAndCloseStream(y);
\end{Verbatim}
\<\EnsuresCalledMethods> is also used for verification of ``wrapper'' types that store resources in fields, to be illustrated in \Cref{sec:class-specification}.  Prior work referred to \<\EnsuresCalledMethods> as   \<@EnsuresCalledMethods>~\cite{KelloggSSE2021}.

\smallskip\noindent
{\bf \<@Owning> and \<\NotOwning>}\label{sec:owning-default}
express a form of lightweight ownership and ownership transfer. When two aliases exist for the
same object, these annotations indicate which alias is responsible for fulfilling
must-call obligations.  Also, an \<@Owning> annotation on a field
declaration indicates that the enclosing class is responsible for
satisfying the field's must-call obligation at the end of its lifecycle
(the ``resource acquisition is initialization'' pattern~\cite[\S
  16.5]{Stroustrup-DesignEvolution}).  See \Cref{sec:class-specification}
for example uses of these annotations.

Unlike full-fledged ownership type systems, as in \citet{ClarkePN98} or the Rust programming language~\cite{klabnik2018rust}, lightweight ownership places \emph{no} restrictions on pointer aliasing in the program.  \<@Owning> and \<\NotOwning> annotations serve only as ``hints'' to the verifier regarding which reference is responsible for closing a resource, and do not impact soundness of verification.  Lightweight ownership suffices since we seek only to verify that resources are freed, not to verify the absence of use-after-free or double-free bugs~\cite{KelloggSSE2021}.
By default, formal parameter types are \<\NotOwning> and return types are \<@Owning>.

\smallskip\noindent
{\bf \<\MustCallAlias>}\footnote{In the implementation, this annotation is
named \<@MustCallAlias>, because it is more general than resources.}
captures a ``resource-aliasing'' relationship. Resource aliases are
either standard must-aliased references \emph{or} references to distinct objects that manage the same underlying resource~\cite{KelloggSSE2021}.  Fulfilling the
must-call obligation of an expression also fulfills the obligation of all of its resource aliases.  \<\MustCallAlias> annotations specify a resource-alias relationship between a method's return value (or, for a constructor, the newly-allocated object) and one of its parameters.
As an example, consider this
method in \<java.net.Socket>:

\begin{Verbatim}[commandchars=\|\[\]]
 class[|bfseries Socket] {
   [|color[blue]|MustCallAlias] OutputStream[|bfseries getOutputStream]([|color[blue]|MustCallAlias] Socket this) { ... }
 }
\end{Verbatim}

The \<\MustCallAlias> annotations denote that the \<OutputStream> returned
by \<getOutputStream> is a resource alias of its receiver argument.  So,
calling \<close()> on the returned \<OutputStream> is equivalent to calling
\<close()> on the \<Socket>: calling either one is sufficient to release
the underlying resource.

\subsection{Resource management specification example}
\label{sec:class-specification}

\begin{figure*}
\begin{smaller}
\begin{minipage}[t]{0.45\textwidth}
\begin{Verbatim}[commandchars=\|\[\],numbers=left,numbersep=.5em]
[|color[blue]@MustCall("dispose")] |label[line:mustCallDisposeClassAnnot]
class[|bfseries MySqlCon] {
  private final [|color[blue]@Owning] Connection[|bfseries con]; |label[line:conField]

  [|color[blue]|MustCallAlias][|bfseries MySqlCon]( |label[line:mustCallAlias1]
     |MustCallAlias Connection[|bfseries con]) { |label[line:mustCallAlias2]
    this.con = con; |label[line:conAssignment]
  }

  void[|bfseries use]() { ... } |label[line:use]

  [|color[blue]@Calls("this.con", "close")] |label[line:disposeCalls]
  void[|bfseries dispose]() { |label[line:dispose]
    closeCon(this.con); |label[line:callCloseCon]
  }

  static [|color[blue]@Owning] Connection[|bfseries createCon]() { |label[line:createCon]
    Connection[|bfseries obj] = ...;
    return obj;
  }
\end{Verbatim}
\end{minipage}
\begin{minipage}[t]{0.45\textwidth}
\begin{Verbatim}[commandchars=\|\[\],numbers=left,numbersep=.5em,firstnumber=21]
  static void[|bfseries useCon]([|color[blue]@NotOwning] Connection obj) |label[line:useCon]
    { ... }

  static void[|bfseries closeCon]([|color[blue]@Owning] Connection obj) { |label[line:closeCon]
    obj.close();
  }
} // end of class MySqlCon


static void[|bfseries client]() { |label[line:client]
  Connection[|bfseries con1] = MySqlCon.createCon(); |label[line:con1]
  MySqlCon[|bfseries mySqlCon1] = new MySqlCon(con1); |label[line:mySqlCon1]
  mySqlCon1.use();
  if (...)
    con1.close();
  else
    mySqlCon1.dispose();
}
\end{Verbatim}
\end{minipage}
\end{smaller}
\caption{Example resource management specifications.
  The Resource Leak Checker can modularly verify the absence of resource
  leaks in this code.
  Given this program without annotations, our resource specification
  inference can infer all the annotations.
}
\label{fig:mot-eg}
\end{figure*}

\Cref{fig:mot-eg} shows a class \<MySqlCon> annotated with a resource management
specification and a client
usage that can be verified as correct.
The
techniques of this paper can automatically infer all the annotations written in the
example, which prior work \cite{KelloggSSE2021} required a human to provide.

In \cref{fig:mot-eg}, field \<con> (line~\ref{line:conField}) has
qualified type \<\MustCall("close")> \<Connection>. It implicitly has the
qualifier \<\MustCall("close")> because \<java.sql.Connection> objects
manage a resource, so the \<Connection> class is annotated as
\<\MustCall("close")> in the JDK (that is, in the verifier's standard
library model).  The \<con> field is annotated as \<@Owning>: this implies
that the enclosing \<MySqlCon> class must have a must-call method that
satisfies \<con>'s must-call obligation.  Accordingly, the \<MySqlCon>
class is annotated \<\MustCall("dispose")>
(line~\ref{line:mustCallDisposeClassAnnot}), and its \<dispose> method is
annotated \<\EnsuresCalledMethods("this.con", "close")>
(line~\ref{line:disposeCalls}), indicating it guarantees \<close()> is
called on \<this.con>.

The \<MySqlCon> constructor stores its argument in field \<con> (line~\ref{line:conAssignment}).
The \<\MustCallAlias> annotations on the constructor (lines \ref{line:mustCallAlias1} and \ref{line:mustCallAlias2}) indicate that the argument
and the new object are ``resource aliases'':
fulfilling the must-call obligations of one
object also fulfills the obligations of the other object.
The Resource Leak Checker can validate method \<client> (line~\ref{line:client}) --- that is, prove that \<client>
releases all resources --- because the \<\MustCallAlias> annotations show that 
\<con1> and \<mySqlCon1> (lines~\ref{line:con1}--\ref{line:mySqlCon1}) are resource aliases.

\<@Owning> and \<\NotOwning> annotations on parameters and return types indicate which object reference is responsible for fulfilling must-call
obligations.  Consider the
static methods in~\cref{fig:mot-eg} (starting at line~\ref{line:createCon}).  The factory method
\<createCon> returns an \<@Owning> \<Connection>, indicating the caller is
responsible for closing it.
\<useCon> has a \<@NotOwning> annotation on its parameter
(line~\ref{line:useCon}), indicating it
will \emph{not} take responsibility for closing its parameter.  \<@Owning>
is the default for returns, and \<@NotOwning> is the default for
parameters, so the annotations on lines~\ref{line:createCon}
and~\ref{line:useCon} are unnecessary; \cref{fig:mot-eg} shows them for emphasis.
\<closeCon> does take ownership of its argument (line~\ref{line:closeCon}), and the \<closeCon> call on
line~\ref{line:callCloseCon} enables the tool to verify the
\<\EnsuresCalledMethods> annotation on line~\ref{line:disposeCalls}.

\subsection{Modular Verification}\label{sec:verification-algorithm}

A program annotated with a resource management specification can be \emph{modularly} verified~\cite{KelloggSSE2021}.  
The key intuition is that resource leak detection 
(but not other related problems, such as proving the absence of use-after-free bugs) 
is an instance of an \emph{accumulation problem}~\cite{KelloggSSE2022}, which is a restricted
class of typestate analysis~\cite{StromY86} that admits sound verification even in the presence of arbitrary, untracked aliasing.
Instead, a verifier can perform an intra-procedural dataflow analysis, computing for each program point a set of pairs $\langle V, e \rangle$, where $e$ is an expression with a non-empty \<@MustCall> obligation, and $V$ is a set of resource-aliased variables referencing $e$.  If a statement $s$ ensures $e$'s \<@MustCall> obligation is satisfied via an operation on some variable in $V$, then $\langle V, e \rangle$ is not propagated to successors of $s$.  If some $\langle V, e \rangle$ reaches a method CFG's exit node, a resource leak warning is reported.

The \<@Owning> and \<\EnsuresCalledMethods> annotations of \cref{sec:annotation-syntax} enable modular verification by informing the verifier when a \<@MustCall> obligation is satisfied via another method or alias.  A \<@MustCall("m")> obligation for $e$ is considered satisfied when \<m> is directly invoked on $e$, but also in the following cases (which rely on annotations): $e$ is returned and the method's return type is \<@Owning>, $e$ is passed to another method's \<@Owning> parameter, $e$ is written to an \<@Owning> field, or $e$ is passed to a method whose \<@Calls> specification ensures \<m> will be called on $e$.  The verifier also uses \<\MustCallAlias> annotations on invoked methods to update its set of resource aliases for an expression, improving precision (e.g., when verifying the \<client> method in \cref{fig:mot-eg}, discussed in \cref{sec:class-specification}).  

A modular verifier must also ensure that all annotations respect standard subtyping rules in the presence of method overriding and other features of real object-oriented languages.  E.g., if a supertype method has a \<\EnsuresCalledMethods> annotation, the \<\EnsuresCalledMethods> annotation on any overriding method must be at least as strong, i.e., it should guarantee that at least the same methods are called.  Similarly, if a supertype method parameter is \<@Owning>, the corresponding parameter in an overriding method must also be \<@Owning>, and if a supertype method return type is \<\NotOwning>, an overriding method's return type should also be \<\NotOwning>.  By enforcing standard subtyping rules, modular verification can proceed \emph{without} constructing a call graph: a call site can be analyzed using the declared target, since any overriding method must respect its specification.

Finally, for soundness, all \<\EnsuresCalledMethods>, \<@Owning>, and \<\MustCallAlias> annotations must be verified, which can also be done modularly.  For further details, see \citet{KelloggSSE2021}.

\section{Inference}
\label{sec:inference-rules}

This section presents declarative rules for our inference algorithm and discusses key properties of the algorithm.  \Cref{sec:optimistic-inference} provides the intuition behind optimistic inference.  \Cref{sec:rules-intro} introduces our inference rule syntax and the base program facts it relies on.  \Cref{sec:phase1} gives rules for phase 1 of inference, and \Cref{sec:phase2} gives rules for phase 2.  Finally, \Cref{sec:inference-key-properties} discusses key properties of the inference algorithm.

\subsection{Optimistic inference}\label{sec:optimistic-inference}

Given a program $P$ (possibly partially annotated), the goal of
our optimistic inference is to discover a resource management specification that most likely matches the intended specification of the developer.  In the case where $P$ is free of leaks, ideally the inferred annotations would
allow a modular verification tool
to verify $P$.
A classical type inference approach is to recast the type-checking rules as
constraints, then solve the constraints.  This approach does not work in
our context because the verification algorithm is not expressible as a type
system (though parts of it are): for example, \<@Owning> is not a type qualifier but instead a hint to the verifier.  Inference in our context therefore requires a
novel algorithm.

Ideally, optimistic inference should infer the programmer's intention even for
\emph{buggy} programs that leak resources on some paths.  Consider a class
\<Socket\-Wrapper> wrapping two sockets \<socket1> and \<socket2>, with the
following \<cleanup()> method:

\begin{Verbatim}[commandchars=\|\[\]]
 void[|bfseries cleanup]() throws IOException {
   socket1.close();
   socket2.close();
 }
\end{Verbatim}

\noindent
This method does not necessarily close \<socket2>, because
\<socket1\-.close()> could throw an exception.   An inference
technique that accurately reflects the code's behavior would infer only
\<\EnsuresCalledMethods("socket1",> \<"close")> for this method, rather than
\<\EnsuresCalledMethods(\{"socket1",> \<"socket2"\},> \<"close")>.
The former specification hinders inference in other parts of the program
and, when used by
a verification tool, leads to confusing false positive alarms at call sites
of \<cleanup>.  By contrast, the latter specification reflects programmer intent, and a
verification tool issues an error within \<cleanup>, exactly where a
programmer needs to fix the bug.  The rules described in the next sections employ this optimistic approach.%

\begin{table*}[t]
  \caption{The input facts that represent program constructs.}
  \centering
  \small
  \setlength\extrarowheight{4pt}
  \begin{tabular}{ |p{4.1cm}||p{9cm}| }
    \hline
    Name &Definition\\
    \hline
    $\InstanceField(f,C)$ & $f$ is a field of class $C$ \\
    $\InstanceMethod(m,C)$ & $m$ is a non-constructor method of class $C$ \\
    $\textsc{AbstractMethod}(m,C)$ & $m$ is an abstract method of class $C$ \\
    $\textsc{Constructor}(m,C)$ & $m$ is a constructor of class $C$ \\
    $\textsc{FieldType}(f,T)$ & field $f$ has basetype $T$ (the basetype elides type qualifiers like \<\MustCall>) \\
    $\textsc{ReturnType}(m,T)$ & method $m$ has return basetype $T$ \\
    $\textsc{ParamType}(m,T)$ & method $m$'s parameter has basetype $T$ \\
    $\textsc{Invokes}(s,m,n,r,p)$ & there \textit{exists} statement $s$ in method $m$ that invokes method $n$ with receiver $r$ and other argument $p$ (optimistic)\\
    $\textsc{WritesField}(s,m,f,v)$ & there \textit{exists} statement $s$ in method $m$ that writes
    variable $v$ to field $f$ of \<this> (optimistic)\\
    $\textsc{NotWrittenAfter}(f,s,m)$ & field $f$ is not assigned
    after statement $s$ in method $m$.  It can be computed from $m$'s control-flow graph. \\
    $\textsc{Returns}(s,m,v)$ & there \textit{exists} statement $s$ in method $m$ that returns $v$ (optimistic)\\
    $\textsc{ThisOrSuperCall}(s,m,m',v)$ & there \textit{exists} statement $s$ in constructor $m$ that is a \<this()> or \<super()> call invoking constructor $m'$, passing $v$ as the parameter (optimistic) \\
    \hline
  \end{tabular}
    \vspace*{-3mm}
\label{tab:inputfacts}
\end{table*}

\subsection{Inference rule syntax and base facts}
\label{sec:rules-intro}

Our inference rule formalism is language-independent, subsuming object-oriented languages
such as C\# and Java.
For simplicity and without loss of generality,
our formalism assumes a name exists for any expression with a non-empty
\<\MustCall> obligation; this can be satisfied  %
via introduction of temporary variables.
In our formalism,
every constructor takes one parameter, and every other method takes two
parameters:  the receiver parameter and one additional formal parameter.
Formal parameters are final (unassignable).
Our formalism does not include static methods and fields.
These restrictions are similar to other
formalisms~\cite{Igarashi:2001:FJM}.
Our implementations handle the full C\# and Java languages.

\Cref{tab:inputfacts} shows the input facts that represent program constructs.
Optimistic inference arises from the fact that our inference uses a
may-analysis, but verification performs a must-analysis.
The optimism originates from the facts in \cref{tab:inputfacts} that
contain ``\emph{exists}''
and propagates through the inference rules shown in
\cref{fig:ecm-and-field-rules,fig:mca-rules}. For instance,
$\textsc{Invokes}(s,m,n,r,p)$ checks whether there exists a statement $s$ in
method $m$ that invokes method $n$ with receiver $r$ and other
parameters $p$. However, verification requires that such a method $n$ is
successfully executed on all paths in $m$.
The inference optimistically assumes that if a programmer wrote a disposal
method call on one path, the programmer intended to write it on all paths.

Our inference rules, presented in \cref{fig:ecm-and-field-rules,fig:mca-rules}, are
written in the style of Datalog rules, though in places they use logical
conditions beyond what Datalog can express (our implementations are not
based on pure Datalog solvers).  The rules are of the form
$\mathit{fact} \leftarrow \mathit{condition},\ldots$,
indicating that the fact is inferred when all conditions after the arrow
hold.  Facts and conditions have parameters that must be matched; the $\_$
parameter always matches.

\sloppypar
In \cref{fig:ecm-and-field-rules,fig:mca-rules}, \textsc{*Annot} facts represent annotations inferred by the
algorithm.  In $\textsc{ParamAnnot}(\_, \_, p)$ (a formal parameter
annotation), $p$ is the name of the method's non-receiver formal parameter.
If the input program already contains a partial resource management
specification, its 
annotations can be
represented as input facts.  

Our inference algorithm proceeds in two phases.
The first phase (\cref{sec:phase1}) infers ownership and disposal
methods:  \<\MustCall> annotations for classes,
\<@Owning> for fields and some method parameters, and 
\<\EnsuresCalledMethods> for disposal methods.
The second phase (\cref{sec:phase2}) infers all remaining annotation types.  Two phases are required since the second phase rules 
rely on the
ownership and disposal method annotations inferred by the first phase.

\begin{figure*}
  \vspace*{-3mm}%
  \small%
  \centering%
    \begin{subfigure}[t]{0.41\textwidth}%
      \begin{align*}
        & \textbf{\textsc{ClassAnnot}}(\<\MustCall($m_\|cd|$)>,C) \leftarrow \quad\Circled{1} &\\
        & \quad \InstanceMethod(m_\|cd|,C), &\\ 
        & \quad \neg \textsc{\textbf{ClassAnnot}}(\<\MustCall($m_\|cd|'$)>,C), &\\
        & \quad \forall f \in \mathit{OwningFields}(C): &\\
        & \quad \quad \textbf{\FieldMustCall}(f,m_\|fd|),&\\
        & \quad \quad \textsc{\textbf{MethodAnnot}}(\<\EnsuresCalledMethods($f$,$m_\|fd|$)>, m_\|cd|) &\\[0.3\baselineskip]
        & \textbf{\textsc{FieldAnnot}}(\<@Owning>,f) \leftarrow \quad\Circled{2} &\\
        & \quad \InstanceField(f,C), \InstanceMethod(m,C), &\\
        & \quad \textbf{\FieldMustCall}(f,m_\|fd|), &\\
        & \quad \textsc{\textbf{MethodAnnot}}(\<\EnsuresCalledMethods($f$,$m_\|fd|$)>,m) &\\[0.3\baselineskip]
        & \textbf{\FieldMustCall}(f,m_\|fd|) \leftarrow \quad\Circled{3} &\\
        & \quad \textsc{FieldType}(f,T), &\\ 
        & \quad \textsc{\textbf{ClassAnnot}}(\<\MustCall($m_\|fd|$)>,T) & \\[0.3\baselineskip]
        & \textbf{\textsc{MethodAnnot}}(\<\EnsuresCalledMethods($f$,$m_\|fd|$)>,m) \leftarrow \quad\Circled{4} &\\
        & \quad \textbf{\FieldMustCall}(f,m_\|fd|), \textsc{Invokes}(s,m,m_\|fd|,f,\_), &\\
        & \quad \textsc{NotWrittenAfter}(f,s,m)%
      \end{align*}
    \end{subfigure}%
    \hfill%
    \begin{subfigure}[t]{0.49\textwidth}
      \begin{align*}
        & \textbf{\textsc{MethodAnnot}}(\<\EnsuresCalledMethods($f$,$m_\|fd|$)>,m) \leftarrow \quad\Circled{5} &\\
        & \quad \textbf{\FieldMustCall}(f,m_\|fd|), &\\ 
        & \quad\textsc{Invokes}(s,m,m',\<this>,\_), &\\
        & \quad \textsc{\textbf{MethodAnnot}}(\<\EnsuresCalledMethods($f$,$m_\|fd|$)>,m'), &\\
        & \quad \textsc{NotWrittenAfter}(f,s,m)\\[0.3\baselineskip]
        & \textbf{\textsc{MethodAnnot}}(\<\EnsuresCalledMethods($f$,$m'$)>,m) \leftarrow \quad\Circled{6} &\\
        & \quad \textsc{Invokes}(s,m,m',\_,f),&\\
        & \quad \textsc{\textbf{ParamAnnot}}(\<@Owning>,m',\_), &\\
        & \quad \textsc{NotWrittenAfter}(f,s,m)\\[0.3\baselineskip]
        & \textbf{\textsc{ParamAnnot}}(\<@Owning>,m,p) \leftarrow \quad\Circled{7} &\\
        & \quad \textsc{ParamType}(m,T), &\\ 
        & \quad \textsc{\textbf{ClassAnnot}}(\<\MustCall($m_\|pd|$)>,T),&\\
        & \quad \textsc{Invokes}(s,m,m_\|pd|,p,\_) &\\[0.3\baselineskip]
        & \textbf{\textsc{ParamAnnot}}(\<@Owning>,m,p) \leftarrow \quad\Circled{8} &\\
        & \quad \textsc{Invokes}(s,m,m',\_,p),  &\\
        & \quad\textsc{\textbf{ParamAnnot}}(\<@Owning>,m',\_)%
      \end{align*}
    \end{subfigure}
    \caption{Phase 1 of inference: rules for inferring \EnsuresCalledMethods, @Owning fields and parameters, and \MustCall\ on
      classes.}
    \label{fig:ecm-and-field-rules}
    \centering
    \begin{subfigure}[t]{0.45\textwidth}
      \centering
      \begin{align*}
        &\textbf{\textsc{AlwaysWrittenToOwningField}}(p,m) \leftarrow \quad\Circled{9} &\\
        & \quad \forall \mathit{path} \in \mathit{NormalPaths}(m).\ \exists\ s \in \mathit{path}. &\\
        & \quad \quad \textsc{\textit{ResourceAlias}}(s,p,r),  \textsc{WritesField}(s,m,f,r),  &\\
        & \quad \quad \textsc{FieldAnnot}(\<@Owning>,f), &\\
        & \quad \quad \textsc{NotWrittenAfter}(f,s,m) &\\[0.3\baselineskip]
        & \textbf{\textsc{ParamAnnot}}(\<@Owning>,m,p) \leftarrow \quad\Circled{10} &\\
        & \quad \textsc{\textit{ResourceAlias}}(s,p,r), \textsc{Invokes}(s,m,m',\_,r), &\\
        & \quad \textsc{\textbf{ParamAnnot}}(\<@Owning>,m',\_) &\\[0.3\baselineskip]
        & \textbf{\textsc{ParamAnnot}}(\<@Owning>,m,p) \leftarrow \quad\Circled{11} &\\
        & \quad \textsc{Constructor}(m,C), |\mathit{OwningFields}(C)| > 1, &\\
        & \quad \textsc{\textbf{AlwaysWrittenToOwningField}}(p,m) &\\[0.3\baselineskip]
        & \textbf{\textsc{ParamAnnot}}(\<@Owning>,m,p) \leftarrow \quad\Circled{12} &\\
        & \quad \textsc{ParamType}(m,T),&\\
        & \quad \textsc{\textbf{ClassAnnot}}(\<\MustCall($m_\|pd|$)>,T),&\\
        & \quad \textsc{\textit{ResourceAlias}}(s,p,r), \textsc{Invokes}(s,m,m_\|pd|,r,\_) %
      \end{align*}
    \end{subfigure}
    \hfill
    \begin{subfigure}[t]{0.45\textwidth}
      \centering
      \begin{align*}
        & \textbf{\textsc{ParamAnnot}}(\<\MustCallAlias>,m,p), \quad\Circled{13} &\\
        & \textbf{\textsc{ReturnAnnot}}(\<\MustCallAlias>,m) \leftarrow &\\
        & \quad \textsc{Constructor}(m,C), |\mathit{OwningFields}(C)| = 1,&\\
        & \quad \textsc{\textbf{AlwaysWrittenToOwningField}}(p,m) &\\[0.3\baselineskip]
        & \textbf{\textsc{ParamAnnot}}(\<\MustCallAlias>,m,p), \quad\Circled{14} &\\
        & \textbf{\textsc{ReturnAnnot}}(\<\MustCallAlias>,m) \leftarrow &\\
        & \quad \textsc{Constructor}(m,C),  \textsc{\textit{ResourceAlias}}(s,p,r), &\\
        & \quad \textsc{ThisOrSuperCall}(s,m,m',r), &\\
        & \quad \textsc{\textbf{ParamAnnot}}(\<\MustCallAlias>,m',\_), &\\
        & \quad \textsc{NotWrittenAfter}(s,f,m) &\\[0.3\baselineskip]
        & \textbf{\textsc{ParamAnnot}}(\<\MustCallAlias>,m,p), \quad\Circled{15} &\\
        & \textbf{\textsc{ReturnAnnot}}(\<\MustCallAlias>,m) \leftarrow &\\
        & \quad \InstanceMethod(m,\_), &\\
        & \quad \forall \mathit{path} \in \mathit{NormalPaths}(m).\ \exists\ s \in \mathit{path}. &\\
        & \quad \quad \textsc{\textit{ResourceAlias}}(s,p,r), \textsc{Returns}(s,m,r)%
      \end{align*}
    \end{subfigure}
    \begin{subfigure}{0.45\textwidth}
      \begin{align*}
        & \textbf{\textsc{ReturnAnnot}}(\<\NotOwning>,m) \leftarrow \quad\Circled{16} &\\
        & \quad \InstanceMethod(m,C), \textsc{ReturnType}(m,T), &\\
        & \quad \textsc{\textbf{ClassAnnot}}(\<\MustCall($\_$)>,T), &\\
        & \quad \InstanceField(f,C), \textsc{Returns}(s,m,f)
      \end{align*}    
    \end{subfigure}
    \caption{Phase 2 of inference: rules for inferring
      \<\MustCallAlias>, \<@Owning>, and \<\NotOwning> parameters.
    }
    \label{fig:mca-rules}
\end{figure*}

\subsection{Phase 1:  ownership and destructors}
\label{sec:phase1}
\Cref{fig:ecm-and-field-rules} gives rules for the
first phase of inference.
$\FieldMustCall(f, m_\|fd|)$ (rule $\Circled{3}$ in \cref{fig:ecm-and-field-rules}) means that $m_\|fd|$ is the direct disposal
method for the type $T$ of field $f$; that is, the declaration of $T$ is
annotated with \<\MustCall($m_\|fd|$)>.
Other methods might also be guaranteed to dispose of $f$, if they are
annotated \<\EnsuresCalledMethods($f, m_\|fd|$)>.

\Cref{fig:ecm-and-field-rules} gives three rules for inferring a
\<\EnsuresCalledMethods>$(f, m_\|fd|)$ method annotation:
when $m_\|fd|$ is invoked directly (rule $\Circled{4}$), when another method $m'$ with an appropriate \<\EnsuresCalledMethods> annotation is invoked (rule $\Circled{5}$), and when $f$ is passed to an \<@Owning> parameter (rule $\Circled{6}$).  For all three cases,
\textsc{NotWrittenAfter} ensures the field is not overwritten after its
\<\MustCall> method is invoked.

Two rules ($\Circled{7}$ and $\Circled{8}$) infer a limited set of \<@Owning> parameters, one for when the
disposal method is directly invoked on a parameter, and one for when a
parameter is passed to another method in an \<@Owning> position.  Rules for
the second phase (\cref{sec:phase2}) infer a larger set of \<@Owning> parameters, but we found that limited inference of \<@Owning> parameters in phase 1 was important for discovering certain class disposal methods.

An \<@Owning> annotation is inferred for a field
when some method in the enclosing class invokes its \<\MustCall> method, as
captured by an inferred \<\EnsuresCalledMethods> annotation (rule $\Circled{2}$).

The rule for inferring \<\MustCall> annotations for a class $C$ (rule $\Circled{1}$) is a bit
more complex.  The rule aims to identify a single instance method $m$ in
$C$ that, when called, is guaranteed to satisfy the \<\MustCall>
obligations of \emph{all} \<@Owning> fields in \<C>.  The rule allows for inferring at most one \<\MustCall> annotation per class.  Multiple suitable methods could be handled by a \<\MustCall> qualifier supporting disjunction, but this is not supported by the current verifiers.

Note that inferring \<\MustCall> on classes relies on inference of \<@Owning> fields, and inference of \<@Owning> fields may rely on inferred \<\MustCall> class annotations.  This cyclic dependence could lead to problems in cases where a class has multiple \<@Owning> fields of a user-defined type, e.g.:
\begin{Verbatim}[commandchars=\|\[\]]
 class[|bfseries Wrapper] {
   [|color[blue]@Owning] MyResource1[|bfseries f1];
   [|color[blue]@Owning] MyResource2[|bfseries f2];
   ...
 }  
\end{Verbatim}

\noindent
Suppose that \<MyResource1> and \<MyResource2> each have a class disposal
method (non-empty \<\MustCall> annotations on the definitions of
\<MyResource1> and \<MyResource2>) that must be discovered by inference.
An issue arises if these \<\MustCall> annotations are discovered at
different times during inference.  In such a case, inference may first
annotate just one of the fields as \<@Owning> and infer a \<\MustCall>
annotation for the class based just on this field.  Then, the discovery of
the second \<@Owning> field may invalidate the previously-inferred
\<\MustCall> annotation.
An inference engine that allowed for retracting inferred facts could handle this scenario.

\label{sec:class-dependencies}
This bad case is rare in practice.
A class
with \<@Owning> fields usually has a \emph{single} public method that
closes all the fields, thereby excluding the possibility of initially
inferring an incorrect \<\MustCall> class annotation.  In general, a
dependence graph between classes could be used to analyze classes in an
order that avoids these ordering issues; we do not formalize this extension.  Our implementations ensure that,
while analyzing class $C$, all \<@Owning> fields within $C$ are discovered
before inferring the \<\MustCall> annotation for $C$ (using the current
\<\MustCall> types for other classes).

\subsection{Phase 2:  remaining annotations}
\label{sec:phase2}

\Cref{fig:mca-rules} gives rules for the second phase of
inference, which handles all remaining annotations.  The \cref{fig:mca-rules} rules for inferring 
\<\MustCallAlias> and \<@Owning> annotations rely on $\textsc{\textit{ResourceAlias}}(s,p,r)$ facts.  The fact $\textsc{\textit{ResourceAlias}}(s,p,r)$ means that variables $p$ and $r$ are resource aliases at the program point immediately before statement $s$.  Resource aliases can be computed via a straightforward extension to any algorithm for computing must-aliased variables.  The three key properties of resource aliases are:
\begin{enumerate}
  \item Every variable is always a resource alias of itself.
  \item All must-aliased pointers at a program point are resource aliases.
  \item Given a call $\mathit{p = m(q)}$, if method $m$ is annotated with \<\MustCallAlias> on its parameter and return type, $p$ and $q$ are resource aliases at the program point immediately after the call.
\end{enumerate}
Due to property 3, resource aliases must be re-computed as new \<\MustCallAlias> annotations are inferred.  Many rules in \cref{fig:mca-rules} allow for operations to be performed through a resource alias of the parameter; below we simply say ``the parameter'' to mean the parameter or its resource aliases.

\<\MustCallAlias> annotations are only valid in pairs, one on the return of
a method and the other on its parameter.  \Cref{fig:mca-rules} has three
rules for inferring these \<\MustCallAlias> pairs, capturing the different
conditions that will allow the annotations to be verified.  The first rule
($\Circled{13}$ in \cref{fig:mca-rules}) infers \<\MustCallAlias> on a constructor if its parameter is always
written into an \<@Owning> field of the class, and if the class has exactly
one \<@Owning> field.  This rule leverages a
helper rule \textsc{AlwaysWrittenToOwningField} (rule $\Circled{9}$), which checks for
appropriate field writes on all normal CFG paths through the method, i.e.,
all paths corresponding to the method exiting without throwing an
exception.  (Our implementations use standard dataflow analysis techniques
rather than enumerating paths.)  The second rule ($\Circled{14}$) captures a constructor
passing its parameter to another constructor that already has a
\<\MustCallAlias> parameter.  The third rule ($\Circled{15}$) handles a method that always
returns (a resource alias of) its parameter.

\Cref{fig:mca-rules} also gives rules for inferring \<@Owning> annotations on parameters.  Phase one of inference inferred a limited number of parameter \<@Owning> annotations (see rules $\Circled{7}$ and $\Circled{8}$ in \cref{fig:ecm-and-field-rules}); in phase two, more can be inferred due to use of resource aliases.  Two rules ($\Circled{10}$ and $\Circled{12}$) match the \textsc{ParamAnnot}(\<@Owning>) rules of \cref{fig:ecm-and-field-rules}, but also allow operations to occur through resource aliases (rule $\Circled{10}$ matches rule $\Circled{8}$, and $\Circled{12}$ matches $\Circled{7}$).  Note that these two rules do \emph{not} require that the operation occur on all paths (that is, they are optimistic); in our experience, an invocation of a \<\MustCall> method strongly implies an intent to take ownership, even if the call does not occur on all paths.  The final rule ($\Circled{11}$) is similar to the first rule for inferring \<\MustCallAlias> (rule $\Circled{13}$) but handles the case where a class has multiple \<@Owning> fields.  In the case of a single \<@Owning> field, we prefer to infer \<\MustCallAlias> on a constructor, since it gives client code the flexibility to finalize either the passed-in resource or the newly-allocated object.

Finally, \cref{fig:mca-rules} gives a rule ($\Circled{16}$) for inferring %
\<\NotOwning> annotations.  
\<\NotOwning> is inferred when a method's return
type has a non-empty \<\MustCall> type and the method acts as a ``getter,''
returning an instance field of the class.  In such cases, verifiers cannot reason about callers satisfying the \<\MustCall> obligation of the field, so there is no purpose in making the return type \<@Owning>.

\subsubsection{Example} We illustrate our inference rules and their
interactions using the example of \cref{fig:mot-eg}.  Assume the program
initially has no annotations.  In phase 1, the first rule from
\cref{fig:ecm-and-field-rules} for \<@Owning> parameters (rule $\Circled{7}$) infers an
\<@Owning> annotation for the \<closeCon> parameter
(line~\ref{line:closeCon}), since \<closeCon> invokes \<close> on its
parameter.  Given this \<@Owning> parameter, the final rule for inferring
\<@Calls> (rule $\Circled{6}$) then applies to \<dispose>, yielding the \<@Calls("this.con",>
\<"close")> annotation on line~\ref{line:disposeCalls}.  In turn, this
\<@Calls> annotation enables the rule for inferring \<@Owning> on fields ($\Circled{2}$),
yielding the \<@Owning> annotation on line~\ref{line:conField}.  Finally,
all annotations are in place to infer \<@MustCall("dispose")> on the
\<MySqlCon> class, via rule $\Circled{1}$ of \cref{fig:ecm-and-field-rules},
concluding phase 1.

In phase 2, \<\MustCallAlias> annotations are inferred for the constructor
of \<MySqlCon> (lines~\ref{line:mustCallAlias1}
and~\ref{line:mustCallAlias2}), via rule $\Circled{13}$ of \cref{fig:mca-rules},
concluding inference for this example.  (Recall that the \<@Owning>
annotation on line~\ref{line:createCon} and the \<@NotOwning> annotation on
line~\ref{line:useCon} are the defaults.)  These inferred annotations
enable the \<client> method in \cref{fig:mot-eg} to pass the verifier.

\subsubsection{Non-final owning fields}
\label{sec:non-final-owning-fields}
The Resource Leak Checker supports
another specification annotation: \<\CreatesMustCallFor(value)>,
which indicates that a method
resets the \<value> expression's must-call obligations.  This annotation
can be useful for specifying certain limited usages of non-final \<@Owning>
fields.  As an example, consider a variant of the \<MySqlCon> class from
\cref{fig:mot-eg}.  In the variant the \<con> field is not final:

\begin{Verbatim}[commandchars=\|\[\]]
  class[|bfseries MySqlCon] {
    private [|color[blue]@Owning] Connection[|bfseries con];
    ... // previous code
    [|color[blue]@CreatesMustCallFor("this")] |label[line:resetCreatesMustCallFor]
    void[|bfseries reset]() {
      if (this.con != null) |label[line:resetConDispose1]
        this.con.dispose(); |label[line:resetConDispose2]
      this.con = createCon();
    }  
  }
  static void[|bfseries client2]() {
    Connection[|bfseries con2] = MySqlCon.createCon();
    MySqlCon[|bfseries mySqlCon2] = new MySqlCon(con2);
    mySqlCon2.use();
    mySqlCon2.reset(); |label[line:resetUse]
    mySqlCon2.use();
    mySqlCon2.dispose();
  }
\end{Verbatim}

\noindent
Here, the \<\CreatesMustCallFor("this")> annotation on \<reset> allows the
Resource Leak Checker to verify that this code is free of leaks.  The
checker ensures that any resource stored in \<con> is disposed before
\<con> is overwritten, and that clients are again obligated to dispose of
\<MySqlCon> objects after \<reset> is called.

We implemented inference of \<\CreatesMustCallFor>, but then disabled it,
due to the fact that only restricted usage patterns can currently be verified.  In our experience, verifiable code using \<\CreatesMustCallFor> like that shown above is rare; most real-world code with non-final fields either uses more complex protocols or is just buggy.  The data in \citet{KelloggSSE2021} itself showed that the overall impact of \<\CreatesMustCallFor> was questionable at best (see Table 3 of \citet{KelloggSSE2021}).

In our experience, inference of \<\CreatesMustCallFor> across large code
bases leads to further problems not described by \citet{KelloggSSE2021}.
For soundness, when a method \<m> is annotated with \<\CreatesMustCallFor>,
the annotation must then also appear on all instance methods of the class
that transitively invoke \<m>, all methods that \<m> overrides, and all
methods that override \<m>.  Further, most clients of these methods with
inferred \<\CreatesMustCallFor> annotations could not be verified, as they
did not follow the restricted usage pattern supported by the verifier.

The stringent rules for verifying \<\CreatesMustCallFor> annotations should
not be surprising, because storing a resource in a non-final \<@Owning>
field is risky: the field could be overwritten, and the only reference to
the resource lost, yielding a leak.  This riskiness has made us believe
that re-assigning non-final \<@Owning> fields is in fact an anti-pattern
that should be avoided whenever possible.  So, our inference is configured
to infer \<@Owning> on non-final fields as appropriate, but \emph{not} to
infer \<\CreatesMustCallFor>.  With this configuration, any overwrite of
such a field will yield a warning from the checker, encouraging the
developer to shift to a different resource management protocol.
\Cref{sec:evaluation} shows that this decision does not significantly
hinder the effectiveness of our inference in practice.

\subsection{Key Algorithm Properties}\label{sec:inference-key-properties}

Regarding termination, our inference algorithm applies the rules of \cref{sec:phase1,sec:phase2} until a fixed
point is reached.  This aspect of the algorithm terminates: the rules only add annotations (never removing them), and there are a finite
number of possible annotations that can be inferred.  (Annotations like
\<\EnsuresCalledMethods> and \<\MustCall> are parameterized, but
there are a finite number of possible parameter values.)  \todo{Someone should look at the remaining text in this para}As noted in \cref{sec:phase2}, \textsc{\textit{ResourceAlias}} facts may need to be re-computed after new \<\MustCallAlias> annotations are inferred.  Assuming resource aliases are computed using a standard dataflow analysis expressible in the monotone framework~\cite{KamUllman77} (e.g., the algorithm given in \citet{KelloggSSE2021}), these re-computations will also terminate, as new \<\MustCallAlias> annotations monotonically increase the set of resource aliases after calls.

The rules in \cref{fig:ecm-and-field-rules,fig:mca-rules} alone do \emph{not} guarantee that inference is deterministic and will produce a unique solution; guaranteeing determinism for our algorithm requires
additional side constraints.
\todo{The following sentence is hard to parse; I cannot determine the
  relationship among its clauses.  Please rewrite.}%
The only possible non-determinism arises from the
rule for inferring \<\MustCall> on classes in
\cref{fig:ecm-and-field-rules} (rule $\Circled{1}$), particularly from cases where there are
multiple candidate \<\MustCall> methods for a class, and for cases with
multiple \<@Owning> fields of user-defined type (previously discussed at the end of \cref{sec:phase1}).  The former issue can
addressed by giving a deterministic rule for choosing the \<\MustCall>
method from possible candidates, and (as noted in \cref{sec:class-dependencies}) the
latter issue can addressed by requiring that classes be processed in order
according to their dependencies, breaking cycles arbitrarily but
deterministically.  Our implementations are deterministic.

Type inference algorithms are often evaluated based on their soundness and
completeness.  A sound type inference algorithm only infers annotations
that are verifiable.  As noted earlier, our optimistic inference is deliberately
unsound, as we found this necessary to best capture the specifications intended
by developers.
However, the combination of optimistic inference plus a sound verification
tool
is sound in the following way: if after inference, the verifier reports no warnings, the
program is guaranteed to be free of resource leaks. This is the same guarantee that the verification tool offers to a human annotator.

A type inference algorithm is complete if it is guaranteed to discover a
set of annotations that would make an input program type check, if such a
set exists.  Unfortunately, there are certain cases where our algorithm is
incomplete.
One case involves types with multiple \<@Owning> fields, like the following:

\begin{Verbatim}[commandchars=\|\[\]]
 class[|bfseries Wrapper] implements Closeable {
   final Socket[|bfseries s1];
   final Socket[|bfseries s2];
  [|bfseries Wrapper](Socket[|bfseries s1], Socket[|bfseries s2]) {
     this.s1 = s1; this.s2 = s2;
   }
   ...
 }
\end{Verbatim}

The problem is a lack of expressivity in the resource management specification
language: there is no way to express that either the \<Wrapper> or
\emph{both} the wrapped \<Socket>s must be closed.
Assuming \<Wrapper> has a \<\MustCall> method that closes both of the fields, our inference will make both of the fields and both the constructor parameters \<@Owning>.  However, consider the following client code (exception handling elided):
\begin{Verbatim}[commandchars=\|\[\]]
 Socket[|bfseries s1] = ...,[|bfseries s2] = ...;
 Wrapper[|bfseries w] = new Wrapper(s1,s2);
 s1.close(); s2.close();  
\end{Verbatim}
Since \<Wrapper>'s constructor parameters were inferred to be \<@Owning>,
client code \emph{must} release those \<Socket>s by invoking
\<Wrapper.close()> to pass the verifier.  The verifier will warn about not
closing \<w> in the code above, even though there is no leak.
We did not encounter code of this type in our experiments. %
If the \<Wrapper> type above had a single \<Socket> field, our algorithm
would infer \<\MustCallAlias> annotations instead, which allow for either the
\<Wrapper> or the wrapped \<Socket> to be closed. 

Similarly, our algorithm may be incomplete if there are multiple valid
\<\MustCall> methods for a class, but the algorithm chooses the wrong one
for the class annotation.  We did observe this to occur rarely in our
benchmarks; see discussion of \cref{fig:missing-anno-eg} in
\cref{sec:evaluation}.  As discussed in \cref{sec:phase1}, a \<\MustCall>
annotation supporting disjunction could solve this issue.

\section{Implementation}\label{sec:implementation}

To demonstrate the generality of our approach, we developed two separate implementations of our inference algorithm. The first
implementation works for Java programs, producing
annotations that are compatible with the original Resource Leak
Checker~\cite{KelloggSSE2021}. The second implementation targets
C\# programs and generates annotations suitable for use with RLC\#~\cite{RLCsharp:2023}, an
independent implementation of resource management verification for C\#.  We describe these implementations in turn.%

\subsection{Java Inference}\label{sec:java-inference}

The Resource Leak Checker~\cite{KelloggSSE2021} is built using the Checker
Framework, a framework for building pluggable type systems and abstract
interpretations (dataflow analyses)~\cite{PapiACPE2008}.  Our Java
inference implementation is also built on the Checker Framework, leveraging
its whole-program inference (WPI) infrastructure~\cite{KelloggDNAE2023,cf-wpi}.  WPI infers
type qualifiers by repeatedly running a checker over the input code, using
facts it derives to insert new qualifiers on each run, until a fixed point
is reached.  This built-in WPI functionality cannot infer resource
management specifications, as the annotations are mostly not type
qualifiers (see \cref{sec:intro}).

Our implementation re-uses the WPI fixed-point infrastructure, running
alongside the original Resource Leak Checker.  After the Resource Leak
Checker runs on each method, inference runs as a post-analysis pass,
applying the rules of \cref{sec:inference-rules} to discover new
annotations.  Inference re-uses intermediate results computed by the
checker, in particular its computation of resource aliases (see
\cref{sec:phase2}); this re-use saves computation and guarantees results
consistent with the checker.  After each iteration, any newly-inferred
annotations are persisted into specification files, and these persisted
annotations are visible to subsequent iterations. In our experiments, the
algorithm converges after an average of six iterations.

Our current implementation is inefficient, in that it re-analyzes all the
program code in each top-level iteration. Using a worklist to only
re-analyze necessary methods and classes would be more efficient, but the
Checker Framework does not support it.  We plan to optimize our
implementation in the future (borrowing techniques from our C\# inference
implementation; see \cref{sec:csharp-inference}).  But, since we expect
inference to be run infrequently, the speed of the current implementation
is not a critical concern.

Our implementation is currently undergoing code review so that it can be incorporated into the Checker Framework. A future release of the framework will include it.

\subsection{C\# Inference}\label{sec:csharp-inference}

RLC\#~\cite{RLCsharp:2023} is a resource leak checker for C\#, built using CodeQL~\cite{codeql}.
Whereas RLC can be viewed as solving an accumulation-based problem,
RLC\# can be viewed as solving a reachability-based problem --- a
significantly different design approach.
RLC\# uses the local data flow engine of
CodeQL~\cite{codeql-csharp-dataflow} for intra-procedural analysis, and it
uses the specification language of \cref{sec:annotation-syntax} for
inter-procedural reasoning (via C\# attributes rather than Java
annotations).

There are two major language-dependent differences between RLC and RLC\#:
\begin{itemize}
\item Java supports the concept of \emph{checked} and \emph{unchecked} exceptions, whereas C\# only has unchecked exceptions.
        Both RLC and RLC\# handle unchecked exceptions unsoundly. This does not impact Java applications because
        all the critical exceptions in Java are checked. However, the impact is significant in C\# applications.
\item For \emph{generic types}, Java supports type erasure which is not supported by C\#. As a result, C\# inference
        must explicitly associate annotations with each bound for the type parameters. This makes adding annotations
        in the source code for generic types difficult. We avoid this issue by adding annotations as logical formulae
        inside the CodeQL query instead of the source code. The logical formula identifies
        the location and the program element in the source where we need to add an annotation.
        Annotations as logical formulae avoids repetitive building of code and creating a CodeQL database
        with every addition of a new annotation.
\end{itemize}

We implemented our inference algorithm using the same infrastructure as that of RLC\#.
The inference rules described in \cref{sec:inference-rules} are expressed as a custom query in the CodeQL query language.  Then, CodeQL manages the inference fixed point computation internally.  As the CodeQL fixed point engine is highly tuned, this inference implementation is much faster than our current Java implementation.  Finally, the CodeQL query generates a CSV file
containing all inferred annotations.  Note that unlike the Java implementation, our C\# inference does not repeatedly run the RLC\# verifier; instead, relevant logic is shared at the CodeQL query level as needed.  This strategy also provides a performance boost, as certain expensive verification computations need not be run during inference.

\begin{table}[t]
	\caption{The portion of hand-written annotations that our algorithm inferred. The ``MWA'' column gives the total number of manually-written annotations for each benchmark.
	}
	\label{tab:percentage-result}
	\begin{tabular}{c|c||@{\,}c@{\,}|c|c|c|c|c|c|c}
		&  & \multicolumn{3}{c|}{\<@Owning>} & \<@Must>- &  \<@Must>- &  &\\ \cline{3-5}
		& & final & non-final &  & \<Call>- & \<Call> &  & \<@Not>- & \\
		& MWA & fields & fields & params & \<Alias> & on class & \<\EnsuresCalledMethods> & \<Owning> & Total
		\\ \hline
		\textbf{\smaller{Service 1}} & 21 &  5/5 & 2/2 & 0/0  & 0/0  & 8/8 &  6/6 &  0/0 & 100\%
		\\
		\textbf{\smaller{Service 2}} & 28 &  3/3 & 5/5 & 1/1  & 2/2  & 8/8 &  8/8 &  1/1 & 100\%%
		\\
		\textbf{\smaller{Service 3}} & 24 &  7/7 & 1/1 & 0/0  & 0/0  & 8/8 &  8/8 &  0/0 & 100\%
		\\
		\textbf{\smaller{Lucene.Net}}& 63 & 7/7 & 7/7 & 13/13  & 6/6  & 13/14 &  12/13 &  2/3  & 95\% %
		\\
		\textbf{\smaller{EF Core}}   & 25 &  1/2 & 3/3 & 0/1  & 2/6  & 5/6 &  5/6 & 1/1 & 68\%
		\\ \hline
		\textbf{\smaller{zookeeper}} & 93 & 6/6 & 12/16 & 3/6  & 18/20  & 12/25 &  7/12 &  8/8 & 71\% %
		\\
		\textbf{\smaller{hadoop-hdfs}}& 91 & 14/17 & 3/3 & 10/12  & 16/23  & 16/18 &  2/7 &  8/11 & 76\% %
		\\
		\textbf{\smaller{Hbase}}     & 35 & 7/7 & 1/1 & 3/3  & 0/2  & 7/11 &  4/6 &  4/5 & 74\%  %
		\\ \hline
		\textbf{\smaller{Total}}  &	-  &  93\% & 89\% & 83\%  & 75\%  & 79\% &  76\% & 83\% & -  
		\\
	\end{tabular}
\end{table}

\section{Evaluation}\label{sec:evaluation}

This section presents an experimental evaluation of our two inference implementations.  Our evaluation aims to answer these research questions:
\begin{itemize}
\item \textbf{RQ1:} How effective is inference in recovering annotations that were previously added manually?
\item \textbf{RQ2:} How effective is inference in exposing true positive
  bugs (resource leaks) related to both library types and user-defined types?
\item \textbf{RQ3:} After running inference, what percentage of the
  verifier warnings relate to missing or incorrect annotations?
\item \textbf{RQ4:} What is the running time of inference?

\end{itemize}
To answer these questions, we ran inference on a suite of Java and C\#
benchmarks.  The results show that our inference technique is effective and
makes the specify-and-check approach for resource management verification
more practical.

\subsection{Recovering Manual Annotations}\label{sec:recovering-manual}

\subsubsection{Benchmarks} To answer RQ1, we collected a suite of Java and C\# benchmarks that had been manually annotated to pass the Resource Leak Checker and RLC\#, respectively.  For Java, we re-used the three large benchmarks from \citet{KelloggSSE2021}, as their artifact provided annotated versions of these benchmarks.  The benchmark sizes are shown in \cref{tab:performance}; we ran on the exact same modules used in \citet{KelloggSSE2021}.  We updated the annotations as needed to work with the most recent version of the Resource Leak Checker.

For C\#, we selected as benchmarks three proprietary microservices (referred to as Service 1, Service 2, and Service 3), and also two open-source projects, Lucene.NET and EF Core.  Lucene.NET~\cite{lucene} is a port of the
Lucene search library to C\#.  EF Core~\cite{efcore} is an object-database mapper that works with a variety of backend databases through a plugin API\@.  
Benchmark sizes are given in \cref{tab:performance}.
We manually added annotations to these benchmarks to provide a baseline for comparison with our inference result.

\subsubsection{Methodology} 

For the Java benchmarks, we utilized the manually annotated version provided by \citet{KelloggSSE2021}. Our inference process does not rely on the presence of manual annotations. Therefore, we removed annotations that were used by the verifier and conducted inference on the unannotated versions of the benchmarks. Subsequently, we calculated the number of manually-written annotations that were successfully identified through the inference process.

For the C\# benchmarks, we performed inference on the unannotated versions of the benchmark. Subsequently, we calculated the number of manually-written annotations that were successfully recovered through the inference process.

\subsubsection{Results} \Cref{tab:percentage-result} shows the percentage of manually-written annotations that were discovered by our inference algorithm, broken down by each type of annotation. 
On average, it recovered 85.5\% of manual annotations, with
73.7\% recovered on average for open-source Java projects,
92.6\% on average for open-source C\# projects,
and 100\% for proprietary C\# microservices.
We
hypothesize that our technique is more effective on the microservices
because those programs were written under a stricter coding discipline,
with more careful review and standards, to prevent leaks in
production services.

Note that there were 54 hand-written \<\CreatesMustCallFor> annotations that are excluded from \cref{tab:percentage-result}, as we found inference to be more effective when inference of \<\CreatesMustCallFor> was disabled, as discussed in \cref{sec:non-final-owning-fields}.  If included, the average percentage of recovered annotations is reduced to 77\% from 85.5\%; 100\% of annotations are still inferred in the proprietary C\# services.

\begin{figure*}[h]
\begin{smaller}
\begin{tabular}{ccc}
\begin{minipage}[t]{0.48\textwidth}
\begin{Verbatim}[commandchars=\|\[\]]
[|color[blue]@MustCall("shutdown")] // hand-written
[|color[blue]@MustCall("closeSockSync")] // inferred
public class[|bfseries Learner] {

  protected [|color[blue]@Owning] Socket[|bfseries sock];

  ...

  [|color[blue]@Calls(value="this.sock", methods="close")]
  void[|bfseries closeSockSync]() {
    try {
      if (sock != null) {
        sock.close();
        sock = null;
      }
      ...
    } catch (IOException e) {
        ...
    }
  }
\end{Verbatim}
\end{minipage}
&
\begin{minipage}[t]{0.48\textwidth}
\begin{Verbatim}[commandchars=\|\[\]]
  [|color[blue]@Calls(value="this.sock", methods="close")]
  void[|bfseries closeSocket]() {
    if (sock != null) {
      if (...) {
        if (closeSocketAsync) {
          ...
        } else {
          closeSockSync();
        }
      }
    }
  }
  
  [|color[blue]@Calls(value="this.sock", methods="close")]
  public void[|bfseries shutdown]() {
    ...
    closeSocket();
    ...
  }
}
\end{Verbatim}
\end{minipage}
\end{tabular}
\end{smaller}
\caption{Simplified example from Zookeeper that shows (1) the complexity of
  inferring the correct \<@MustCall> annotation, (2) the benefits of our
  optimistic analysis to infer correct annotations that captures
  programmers' intention.\todo{How does the figure show (2)?}}
\label{fig:missing-anno-eg}
\end{figure*}

Here we present some examples from the Java benchmarks where our inference
failed to infer a handwritten annotation.  \Cref{fig:missing-anno-eg} shows
a simplified example from Zookeeper where inference missed
\<@MustCall("shutdown")> on the \<Learner> class.  Three methods of the \<Learner> class
satisfy the constraints defined in \cref{sec:rules-intro} for being a
disposal method for a class. Without information on how \<Learner>
instances are used, it is difficult to determine which method is the true
disposal method.  In this case, inference added \<@MustCall("closeSockSync")> to
the class instead of the desired \<@MustCall("shutdown")>.  We believe that
a better design for this class would have made the \<closeSockSync> and
\<closeSocket> methods private, in which case inference would have added
the correct annotation.

\begin{figure*}[h]
\begin{smaller}
\begin{tabular}{ccc}
\begin{minipage}[t]{0.47\textwidth}
\begin{Verbatim}[commandchars=\|\[\]]
[|color[blue]@MustCall("compute")] // not inferred
abstract static class[|bfseries BlockChecksumComputer] {

  // @Owning annotations that were 
  // not inferred by our analysis
  private final [|color[blue]@Owning] 
     LengthInputStream[|bfseries metadataIn];
  private final [|color[blue]@Owning] 
     DataInputStream[|bfseries checksumIn];
  ...
  
  [|color[blue]@NotOwning]
  LengthInputStream[|bfseries getMetadataIn]() {
      return metadataIn;
  }

  [|color[blue]@NotOwning]
  DataInputStream[|bfseries getChecksumIn]() {
    return checksumIn;
  }
  
  [|color[blue]@Calls(]
    [|color[blue]value={"this.checksumIn", "this.metadataIn"},]
    [|color[blue]methods={"close"})] // not inferred
  abstract void[|bfseries compute]() throws IOException;
}
\end{Verbatim}
\end{minipage}
&
\begin{minipage}[t]{0.5\textwidth}
\begin{Verbatim}[commandchars=\|\[\]]
class[|bfseries ReplicatedBlockChecksumComputer]
  extends BlockChecksumComputer {

  ...
  
  [|color[blue]@Calls(]
    [|color[blue]value={"this.checksumIn", "this.metadataIn"},]
    [|color[blue]methods={"close"})] // not inferred
  void[|bfseries compute]() throws IOException {
    try {
      ...
    } finally {
      IOUtils.closeStream(getChecksumIn());
      IOUtils.closeStream(getMetadataIn());
    }
  }
}
\end{Verbatim}
\end{minipage}
\end{tabular}
\end{smaller}
\caption{Simplified example from Hadoop to illustrate missed annotations and an anti-pattern.}
\label{fig:missing-anno-eg2}
\end{figure*}

\Cref{fig:missing-anno-eg2} gives another example where our inference fails to
infer manually-written annotations.  The
\<BlockChecksumComputer> class from Hadoop contains two private resource
fields. However, its disposal method is defined as abstract, and delegates
the responsibility of closing these resources to the sub-classes that do
not have direct access to the variables. This design choice can be
problematic not only for the modular verifiers and inference but also for
developers, as it can introduce leaks if programmers do not release the
resources properly.  Our inference does not detect patterns where the
resources for \<@Owning> fields are only released in subclasses, so it
misses the the \<@Calls> annotation on \<compute>, the \<@Owning>
annotations on the fields, and the \<@MustCall("compute")> annotation on
the class.

\subsection{Impact on Verifier Warnings}\label{sec:checker-errors}

\begin{table}[t]
\caption{Inferred resource management specifications and causes for checker
  warnings.
A false positive is correct code that the verifier cannot prove safe, even
after annotations are added; the table categorizes this separately from
checker warnings that can be eliminated by adding an annotation.
The sum of all percentages in each row adds to 100\%.
``@CMCF'' indicates the percentage of warnings reported due to missing \<\CreatesMustCallFor> annotations for non-final/non-readonly fields. ``@MCE'' (MustCall Empty) represents the percentage of warnings reported due to missing \<\MustCall()> annotations on class declarations or type uses that do not retain a resource.}
         \label{tab:rlc-inference-eval}
\begin{tabular}{c||@{\,}c@{\,}|c||c|c|c|c|c|c}
& \begin{tabular}{@{}c@{}} \#warnings \\ (no annos) \end{tabular}
& \begin{tabular}{@{}c@{}} \#warnings \\ (inferred \end{tabular}
& \begin{tabular}{@{}c@{}} true \\ posi- \end{tabular}
& \begin{tabular}{@{}c@{}} false \\ posi- \end{tabular}
& \begin{tabular}{@{}c@{}} incorrect \\ annota- \end{tabular}
& \multicolumn{3}{c}{\begin{tabular}{@{}c@{}} missing \\ annotations \end{tabular}}
\\ \cline{7-9}
    &  & annos) & tives & tives & tions & @CMCF & @MCE & Other
\\ \hline
    \textbf{\smaller{Service 1}}   & 251 & 240 & 12\% & 43\% & 0\% &    8\% &  32\% &   5\%
\\
    \textbf{\smaller{Service 2}}   &  45 & 34 & 29\% & 36\% & 0\% &    3\% &  23\% &   9\%
\\
    \textbf{\smaller{Service 3}}   &  20 & 12 & 50\% & 50\% & 0\% &    0\% &   0\% &   0\%
\\
    \textbf{\smaller{Lucene.Net}}  & 670 & 592 & 30\% & 42\% & 2\% &    3\% &  20\% &   3\%
\\
    \textbf{\smaller{EF Core}}     &  88 & 154 & 22\% & 60\% & 0\% &    8\% &   5\% &   5\%
\\ \hline
    \textbf{\smaller{zookeeper}}   & 138 & 170 & 19\% & 49\% & 5\% & 13.5\% &   6\% & 7.5\%
\\
    \textbf{\smaller{hadoop-hdfs}} &  26 & 95 & 18\% & 56\% & 7\% &  9.5\% & 9.5\% &   0\%
\\
    \textbf{\smaller{Hbase}}       & 828 & 844 & 19\% & 44\% & 2\% &    7\% &   9\% &  19\%

\\
\end{tabular}
\label{tab:rlc-inference-eval-java}
\end{table}

For RQ2 and RQ3, we ran the verifier on two versions of each benchmark, first with
no annotations, and second with the annotations inferred by our algorithm.
Then, for the checker warnings reported after inferring annotations, we
categorized them by whether they were caused by incorrect
inferred annotations, missed annotations, resource leaks (true positive bugs), or false
positive warnings from the checker. A true positive is a real resource leak, while a false positive is correct code that the verifier cannot prove safe, even after annotations are added.  As there were too many warnings to triage all of them, we randomly chose at least 50 warnings from each benchmark (or all
warnings if the total number of warnings is less than 50) to
categorize. \Cref{tab:rlc-inference-eval} shows the results.

\begin{figure*}
\begin{smaller}
\begin{tabular}{ccc}
\begin{minipage}[t]{0.48\textwidth}
\begin{Verbatim}[commandchars=\|\[\]]

public class[|bfseries ConnectionWrapper] {
  private final Connection[|bfseries con];
  public[|bfseries ConnectionWrapper]() {
    [|color[red]// warning: assign to non-@Owning field]
    this.con = new Connection(...);
  }

  public void[|bfseries close]() {
    this.con.close();
  }
}
...
[|color[red]// no warnings here]
ConnectionWrapper[|bfseries cw] = new ConnectionWrapper();
ConnectionWrapper[|bfseries cw2] = new ConnectionWrapper();
...
... no calls to close ...
\end{Verbatim}
\end{minipage}
&
\begin{minipage}[t]{0.48\textwidth}
\begin{Verbatim}[commandchars=\|\[\]]
[|color[blue]@MustCall("close")]
public class[|bfseries ConnectionWrapper] {
  private final [|color[blue]@Owning] Connection[|bfseries con];
  public[|bfseries ConnectionWrapper]() {
    [|color[red]// no warnings here]
    this.con = new Connection(...);
  }
  [|color[blue]@Calls(value="this.con", methods="close")]
  public void[|bfseries close]() {
    this.con.close();
  }
}
...
[|color[red]// two warnings: disposal method not called]
ConnectionWrapper[|bfseries cw] = new ConnectionWrapper();
ConnectionWrapper[|bfseries cw2] = new ConnectionWrapper();
...
... no calls to close ...
\end{Verbatim}
\end{minipage}
\end{tabular}
\end{smaller}
\caption{Location of verifier warnings before (left) and after (right) inference.}
\label{fig:loc-of-warning-using-inference}
\end{figure*}

\subsubsection*{Number of Warnings} Per \cref{tab:rlc-inference-eval-java},
inference sometimes increases the number of warnings reported by the
verifier, compared to an unannotated program.
For resource leaks, this result is expected, since our inference
discovers new \<@MustCall> obligations that the verifier does not check if
the code is unannotated. Consider the simple example in
\cref{fig:loc-of-warning-using-inference}, where instance field \<con>
contains a Java \<Connection>.  For unannotated code (left), the
checker reports a single warning that this object is being written into
a non-\<@Owning> field.  With inferred annotations (right), \<con>'s inferred
\<@Owning> annotation leads to inferring a \<@MustCall> obligation for the
\<ConnectionWrapper> class.  If in multiple places, client code uses
\<ConnectionWrapper> objects without calling its disposal method, multiple warnings
are then reported, an increase over the unannotated program.  But, the
warnings after inference are of a higher quality, since they better reflect
the intended resource management specification of the program.  For this
reason, we focus our evaluation on assessing the quality of warnings reported
after inference.

\subsubsection*{Impact of Optimistic Inference} During manual
triage, we found multiple cases where optimistic inference provided better
results than a technique that only infers verifiable annotations.
Consider the \<shutdown> method in \cref{fig:missing-anno-eg}, which serves as the disposal method of the class and calls \<closeSocket>. There are some paths in \<closeSocket> in which the \<sock> object is either null or already closed, and hence \<close> is not called.  However, our algorithm still optimistically annotates the \<closeSocket> method with the \<@Calls("this.sock",> \<"close")> annotation, matching the code's intention to guarantee that the \<Socket> is closed. This annotation is subsequently used by the checker to verify the \<@Calls("this.sock",> \<"close")> annotation on the \<shutdown> method.

\begin{figure*}
\begin{smaller}
\begin{Verbatim}[commandchars=\|\[\]]
[|color[blue]@Calls(value = {"this.storage", "this.committedTxnId", "this.curSegment"}, methods = {"close"})]
public void[|bfseries close]() throws IOException {
  storage.close();
  // committedTxnId remains open if storage.close() throws an exception
  IOUtils.closeStream(committedTxnId);
  IOUtils.closeStream(curSegment);
}
\end{Verbatim}
\end{smaller}
\caption{Simplified example from Hadoop that shows the benefit of optimistic inference.%
}
\label{fig:missing-anno-eg3}
\end{figure*}

Another example is shown in \cref{fig:missing-anno-eg3}, where \<committedTxnId> and \<curSegment> point to resource objects that remain open on the possible exceptional exit caused by the \<storage.close()> call. However, our inference algorithm is able to infer the \<@Calls(\{"this.storage", "this.curSegment",> \<"this.committedTxnId"\},> \<\{"close"\})> annotation.
 This annotation leads to an error report within the \<close()> method, the location that actually requires a fix.
 
\subsubsection*{True and False Positives}\label{sec:bug-finding}
As shown in \cref{tab:rlc-inference-eval-java}, the verifier reports an average of 28\% and 19\%
(out of total warnings) true positives in C\# and Java benchmarks
respectively after inference runs.  This true positive rate is very close
to the average 26\% precision reported for the Resource Leak
Checker~\cite{KelloggSSE2021}, where precision is the ratio of true positive warnings to all tool warnings. However, that work reported precision \emph{after}
laborious work to manually annotate the programs.  The
fact that we achieve a similar precision rate shows that our inference is
highly effective.  Also, \citet{KelloggSSE2021} reports
statistics only about library types, due to the voluminous number of errors
reported and the difficulties of manual annotation.  Our statistics cover
(a random sample of) all errors: both library and user-defined types. While
investigating the random sample, we discovered 6 true positive leaks of
user-defined resources in Java projects, which had been ingored in the
previous work. We also discovered 10 more true positives for user-defined
types in C\# benchmarks that were not reported by RLC\# with manual
annotations.

The Resource Leak Checker and RLC\# report a significant number of false positives on our benchmarks, 46\% and 48\% of the total warnings for C\# and Java benchmarks respectively.  These tools have a high false positive rate because they are sound and do not use heuristics to filter warnings.  Still, previous work showed that the false-positive rate from the RLC was comparable to that of heuristic (unsound) checkers~\cite{KelloggSSE2021}.  
The two major reasons we observed for false positives were
\begin{inparaenum}[(a)]
\item conservative handling of collection types such as lists and
  dictionaries (i.e., poor handling of generics), and
\item a lack of path sensitivity in computing must-alias information, resulting in over-approximation.
\end{inparaenum}

\subsection{Redundant/Incorrect and Missing Annotations} 
\Cref{tab:rlc-inference-eval-java} shows that our technique infers very few incorrect annotations. \Cref{fig:missing-anno-eg} gave an example of an incorrect annotation being inferred (the \<@MustCall("closeSockSync")> annotation on the class).  Overall, the low rate of incorrect inferred annotations shows that our optimistic technique usually infers annotations matching the intended specification.

\Cref{tab:rlc-inference-eval} shows that an average of 25\% and 27\% of the verifier warnings are generated
because of missing annotations across C\# and Java benchmarks respectively.

In C\#, nearly 16\% of warnings are generated because of a missing
\<\MustCall("")> annotation.  We found that in C\#, classes often implement
the \<System.IDisposable> interface, even though they do not manage a
resource that needs to be disposed. In these cases, the \texttt{Dispose} method is used to reset the state of the instance variable and not to dispose any resources. Since we annotate the \<IDisposable>
type as \<@MustCall("Dispose")>, by default this leads to false warnings
when objects of these classes do not call \<Dispose>.  This issue can be
addressed by annotating such classes as \<@MustCall("")>, but our inference
technique cannot yet infer this annotation; we leave this enhancement as
future work.

For Java benchmarks, nearly 10\% of warnings are generated due to
re-assignment of non-final \<@Owning> fields.  As discussed in
\cref{sec:non-final-owning-fields}, such cases are typically code smells.
They could be addressed by adding a \<\CreatesMustCallFor> annotation, but
we found that exhaustive inference of these annotations was not effective.
Our tool's current approach, which leads to verifier warnings
any time a non-final \<@Owning> field is overwritten,
encourages a cleaner programming style for resources.

\subsection{Run-time Performance}\label{sec:performance}

\begin{table}[t]
    \centering
\caption{Time to inferring resource management specifications.
  ``kLoC'' is non-comment, non-blank lines of code.
\todo{I find the use of minutes and seconds really hard to read.  I suggest
  that we report decimal minutes (with a fractional part).}
        }
        \label{tab:performance}
\centering
\begin{tabular}{c|r|r|r|r|r}
    &  & \multicolumn{1}{c|}{Inference} & \multicolumn{3}{c}{Verification time} \\
    & \multicolumn{1}{c|}{kLoC} & \multicolumn{1}{c|}{time} &
  \multicolumn{1}{c|}{no annos} & \multicolumn{1}{c|}{inferred annos} &
  \multicolumn{1}{c}{manually written annos} \\
    \hline
    \textbf{Service 1} & 450 & \zph 2m 37s & \zzph 2m 18s & 3m 38s & 3m 54s
    \\
    \textbf{Service 2} & 163 & \zph 4m 48s & \zzph 3m 31s & 6m 40s & 6m 44s
    \\
    \textbf{Service 3} & 147 & \zph 4m 47s & \zzph 3m \zph 0s & 5m \zph 1s & 5m \zph 9s
    \\
    \textbf{Lucene.Net} & 412 & 11m 17s & \zph 55m 56s & 58m 48s & 57m 48s
    \\
    \textbf{EF Core} & 233 & 10m 53s & 76m 27s  & 95m 36s & 77m 48s
    \\ \hline
    \textbf{zookeeper} & \zph 45 & \zph 6m 16s & 0m 33s & 0m 38s & 0m 28s
    \\ 
    \textbf{hadoop-hdfs} & 152 & 55m 23s & 5m 48s & 4m 37s & 3m 22s
    \\ 
    \textbf{hbase} & 221 & 25m 48s & 2m 52s & 7m 59s & 6m 14s
  \end{tabular}
\end{table}

\Cref{tab:performance} gives performance results for inference and checking.
For C\#, we ran the inference algorithm and measured the run time as an average of 3 trials
on a machine with an Intel Xeon(R) W-2145 CPU running at 3.7 GHz and 64 GB of RAM\@. For Java, we ran the inference algorithm and the Resource Leak Checker on a machine with a 12th Gen Intel Core i-7-12700 Processor, which has 20 cores, and 32 GB of RAM\@.%

As noted in \cref{sec:java-inference}, our Java implementation suffers from some inefficiencies, which are reflected in the numbers.  The CodeQL-based inference implementation for C\# is quite performant, always running in under 12 minutes for benchmarks with hundreds of thousands of lines of code.  We believe with a more optimized architecture we could build an equally-performant inference implementation for Java.  Note that for C\#, inference actually runs faster than running the RLC\# checker.  This difference is due to the RLC\# verifier needing to verify properties on all paths through a method, while optimistic inference often only checks for the existence of an operation on some path.  In certain cases, verification time was higher with the inferred annotations than with manual annotations; this occurred because inference discovered user-defined types that managed resources (recall that such types were not inspected during manual annotation), and the verifier was then obligated to check usages of these types for leaks.

\subsection{Threats to Validity}\label{sec:threats}

For external validity, the primary threat for our evaluation is our choice
of benchmarks.  For Java, we chose benchmarks from previous
work~\cite{KelloggSSE2021}.  For C\#, we strove to choose
representative benchmarks.  However, our inference may perform
differently on other types of benchmarks or coding patterns.

Our results may also be impacted by implementation bugs, threatening
internal validity.  We have a suite of regression tests designed to detect
such bugs, and we have also done extensive manual inspection of the output
of the inference implementations on our benchmarks.
\todo{Mention how much of our implementations are publicly available, to
  permit independent examination.}

\section{Related Work}\label{sec:related}

\paragraph{Static Resource Leak Detection} Several static analyses have
been designed to detect resource leaks in unannotated code.
Tracker~\cite{TorlakC10} and Grapple~\cite{zuo2019grapple} both employ
inter-procedural dataflow analysis to detect resource leaks for Java.
InferSharp~\cite{infer-sharp}, built on Facebook
Infer~\cite{CalcagnoDDGHLOPPR2015}, also leverages inter-procedural analysis to
detect resource leaks for C\# code.  Other tools use more heuristic
approaches and intra-procedural analysis to detect leaks, e.g., analyses in
Eclipse~\cite{ecj-resource-leak} and PMD~\cite{pmd-resource-leak}.

\citet{KelloggSSE2021} directly compared the Resource Leak Checker to the Grapple~\cite{zuo2019grapple} and Eclipse~\cite{ecj-resource-leak} tools.  Compared to Grapple, the Resource Leak Checker had many fewer false negatives and ran more quickly.  The Eclipse checker ran more quickly than the Resource Leak Checker, but 
suffered from false negatives.  The key disadvantage of the Resource Leak Checker was that unlike the other two tools, the benchmarks had to be manually annotated before it could be used.  The inference technique presented in this paper significantly decreases this need for manual annotation, making use of the Resource Leak Checker much easier and more compelling.

\paragraph{Inference} Type inference is a well-studied technique; for
general background see Pierce~\cite{pierce:2002:types-and-pls}.  Most type
inference techniques either infer a set of types for a program that allow
it to pass the type checker, or fail with an error message.  Our scenario
is quite different, as our inference must produce a useful partial solution
for programs that cannot pass the type checker (due to bugs or due to code
patterns that cannot be verified).  Some recent techniques leverage
constraints and optimization for type inference, e.g., approaches to
migrate existing programs to use a gradual type
system~\cite{PhippsCostinAGG2021,MigeedP2020,CamporaCEW2018}.
These techniques can output a partial typing when type inference cannot
fully succeed.  However, the annotations required by the Resource Leak
Checker are not typical type qualifiers, but instead capture ownership and
aliasing protocols that reveal which code is responsible for disposing of a
resource.  The nature of these annotations necessitates a custom algorithm.

\citet{vogels2011annotation} presents an inference technique for separation
logic. This approach is similar to ours in that their inference is
implemented separately from their verifier, thus the inference cannot
introduce unsoundness.  However, there are significant differences in the
inference technique: e.g., their work guesses many annotations and uses the
verifier to see which are valid (like the Houdini
technique~\cite{FlanaganL2001:Houdini}), whereas our technique uses program
analysis to discover likely specifications.  Also, the properties being
checked and the scale of programs being analyzed are notably different.

\citet{HackettDWY2006} presents a modular checker for buffer overflows that
includes an inference approach similar in spirit to ours.  The annotations
for buffer overflow checking require a customized inference algorithm, as
in our case for resource leaks.  Their algorithm is also formulated using
Datalog rules.  And, like our technique, for practicality their algorithm
is optimistic and may infer annotations that cannot be verified.
We believe that this optimistic approach for inference is more general than
buffer overflows and resource leak specifications, and it may be useful in
other domains.

Recent work has applied machine learning to type
inference~\cite{HellendoornBBA2018,PradelGLC2020,PengGLGLZL2022}.  These
approaches again focus on inference of traditional types for variables, and
hence cannot be directly applied to inferring the kinds of annotations
required by the Resource Leak Checker.  Further, applying ML techniques to
our inference problem could be challenging due to a lack of training data.
Still, ML techniques could be complementary to our approach and help to
infer better annotations in cases where our algorithm is incomplete or to
tune its heuristics.

\section{Conclusions}

We have presented a novel technique for inference of resource management
specifications, which enables broader usage of specify-and-verify tools for
modular verification that no resources are leaked.  Our technique leverages
a custom algorithm for handling inter-related specifications of ownership,
aliasing, and resource obligations.  Our technique employs optimistic inference
to more often capture the intended specification for code, even when the code
cannot be verified.  Our experimental evaluation showed that our technique to be
very effective: it inferred most of the annotations developers wrote by
hand, and the final true positive rate of the checker run after
fully-automatic inference nearly matched the rate achieved after manual
annotation.

\begin{acks}
We thank the anonymous reviewers for their detailed and helpful feedback.
This research was supported in part by the National Science Foundation
under grants
2005889,
2007024,
2312262, and
2312263,
DARPA contract
FA8750-20-C-0226, a gift from Oracle Labs, and a Google Research Award.
\end{acks}
%

%

%

\end{document}